\documentclass{article}

\usepackage{arxiv}

\usepackage[utf8]{inputenc} 
\usepackage[T1]{fontenc}    
\usepackage{hyperref}       
\usepackage{url}            
\usepackage{booktabs}       
\usepackage{amsfonts}       
\usepackage{nicefrac}       
\usepackage{microtype}      
\usepackage{graphicx}
\usepackage{doi}
\usepackage{amsmath}
\hypersetup{
    colorlinks,
    linkcolor=black,
    citecolor=black,
    urlcolor=black
}

\title{Second-harmonic optical diffraction tomography}

\author{Amirhossein Saba, Carlo Gigli, Ye Pu, and Demetri Psaltis.\\
School of engineering, Ecole Polytechnique Fédérale de Lausanne, \\
Lausanne, 1015, Switzerland \\
	\texttt{carlo.gigli@epfl.ch} \\
}

\renewcommand{\headeright}{ }
\renewcommand{\undertitle}{ }
\renewcommand{\shorttitle}{Second-harmonic optical diffraction tomography}

\hypersetup{
pdftitle={Second-harmonic optical diffraction tomography},
pdfsubject={q-bio.NC, q-bio.QM},
pdfauthor={A. Saba},
pdfkeywords={First keyword, Second keyword, More},
}

\begin{document}
\maketitle

\begin{abstract}
Optical diffraction tomography (ODT) has emerged as an important label-free tool in biomedicine to measure the three-dimensional (3D) structure of a biological sample. In this paper, we describe ODT using second-harmonic generation (SHG) which is a coherent nonlinear optical process with a strict symmetry selectivity and has several advantages over traditional fluorescence methods. We report the tomographic retrieval of the 3D second-order nonlinear optical susceptibility using two-dimensional holographic measurements of the SHG fields at different illumination angles and polarization states. The method is a generalization of the conventional linear ODT to the nonlinear scenario. We demonstrate the method with a numerically simulated nanoparticle distribution and an experiment with muscle tissue fibers.  Our results show that SHG ODT does not only provide an effective contrast mechanism for label-free imaging but also due to the symmetry requirement enables the visualization of properties that are not otherwise accessible. 
\end{abstract}


\section{Introduction}\label{sec:1}
Optical diffraction tomography (ODT) is a three-dimensional (3D) imaging method that uses multiple 2D quantitative phase images (amplitude and phase) to calculate the 3D refractive index distribution of an inhomogeneous sample. The sample is illuminated with plane waves at different angles. and the complex scattered fields are measured using off-axis holography. ODT is based on the Born approximation and is known as the Fourier diffraction theorem or the Wolf transform. The method was proposed by Emil Wolf \cite{wolf1969three} in 1969, but was experimentally demonstrated 40 years later \cite{sung2009optical}. Thereafter, ODT has been extensively studied for different biological applications \cite{jin2017tomographic, park2018quantitative}, and iterative solutions for ODT were 
also proposed by different groups \cite{lim2015comparative, kamilov2015learning, tian20153d, saba2022physics}. While most of the ODT techniques were limited to scalar refractive index, recently polarization-sensitive ODT methods were proposed to reconstruct the 3D distribution of the dielectric tensor \cite{saba2021polarization, shin2022tomographic}.

Second-harmonic generation (SHG), among other nonlinear optical harmonic generation processes, provides a mechanism of contrast as the coherent counterparts to fluorescence \cite{boyd2020nonlinear}. When an intense optical wave interacts with certain materials, two photons can be annihilated with the simultaneous generation of a photon at a double frequency. The probability for this to happen depends on the intrinsic second-order nonlinear optical susceptibility  $\chi^{(2)}$, which is a tensor, and the phase matching condition due to material spectral dispersion. The process of SHG is highly selective in material symmetry, being only efficient in a material lacking an inversion center in molecular structure. The second-order susceptibility vanishes in all centrosymmetric and isotropic materials. In this way, a sharp contrast is created based on the symmetry of the underlying molecular arrangement of the material of interest. Since the first demonstration of SHG imaging in biological samples with conventional laser-scanning microscopy of rattail tendon using tightly focused ultrashort laser pulses \cite{freund1986second}, SHG microscopy as a useful label-free and background-free imaging modality has received many applications in biomedicine \cite{freund1986second,stoller2003quantitative, plotnikov2006characterization, nucciotti2010probing, dubreuil2018polarization}. By limiting the imaging volume to one small voxel at a time at the focus of the incident light and scanning the 3D sample pot-by-point, such an approach reveals rich structural information from the spatial variation of the SHG intensity. However, quantitative determination of the spatial distribution of the second-order nonlinear susceptibility, $\chi^{(2)}$, is often hindered. In conventional, scalar, ODT the refractive index measurement lacks molecular or structural specificity that is often required in biomedical research. Therefore, SHG can extend the applicability of ODT by providing a richer contrast mechanism through the tensorial measurements. In comparison with fluorescence, SHG has several advantages. The emission of SHG is highly stable since it does not suffer from blinking or photobleaching. SHG  is a narrow-band signal, spectrally far from the fundamental wavelength, making it easier and more efficient to remove the background scattering and emissions. The lossless SHG process itself deposits no photoenergy in the sample and results in little phototoxicity and photodamage, however, photodamage can occur due to the high intensity of the pulsed excitation. SHG also provides great flexibility in the choice of excitation and emission spectral band. Finally, the coherent emission in SHG allows interferometric techniques such as holography.

In recent years, wide-field coherent imaging techniques such as quantitative phase imaging (QPI) and ODT have gained substantial momentum. Given the coherence property of SHG, a combination of the above techniques with SHG would lead to highly useful new imaging modalities with capabilities previously unattainable. Indeed, Pu, et al.  \cite{pu2008harmonic} have demonstrated holography using SHG signals, which was termed harmonic holography. Masihzadeh et al. \cite{masihzadeh2010label} experimented with harmonic holography in biological specimens, showing a 3D collagen structure based on the reconstructed SHG intensity or field. Schaffer et al. \cite{shaffer2011second} also presented SHG QPI in collagen fibers using harmonic holography. In pursuing a 3D technique using SHG signals, Hu et al. \cite{hu2020harmonic} performed the first experiment in a scalar formalism assuming undepleted pump and Born approximation, not in the ODT configuration, but using short-coherent-gated 3D sectioning and axial scanning which has a limited axial resolution. Additionally, due to the tensorial nature of $\chi^{(2)}$, this scalar treatment is inadequate in revealing the true properties of the underlying nonlinear optical structure. More recently, a fully tensorial theoretical formalism was developed for SH-ODT \cite{yu2022optical} based on Wolf’s transform for the 3D reconstruction of $\chi^{(2)}$, which, however, was not experimentally validated.

In this paper, we present for the first time experimental tomographic reconstructions of second-order susceptibility using our polarization-sensitive ODT formalism. In addition, we demonstrate second-harmonic optical diffraction tomography (SH-ODT) for the reconstruction of the 3D distribution of the $\chi^{(2)}$ of the sample. The 3D tomographic reconstruction is based on the 2D complex SH-generated fields by the sample, called SH projections, which are measured using harmonic holography as the sample is illuminated with a high-power pulsed laser from different angles. Polarization-sensitive ODT that we presented in our previous work \cite{saba2021polarization} was the first step towards the generalization of ODT for the 3D reconstruction of properties rather than scalar refractive index distribution. To reconstruct the 3D distribution of the second-order susceptibility in an ODT configuration using Wolf’s transform, a similar approach to our previous study of polarization-sensitive ODT can be used. A similar theoretical formalism was described in \cite{yu2022optical} for SH-ODT. In this paper, we also extend the SH-ODT methodology using regularized iterative reconstruction. We provide synthetic data for digital $\chi^{(2)}$ phantoms to validate the formalism and we demonstrate our methodology by reconstructing the 3D Barium-titanate (BTO) nano-particles and muscle tissue samples, experimentally.


\section{Results}
\label{sec:2}
\subsection{Principle}
\begin{figure}[t]
    \centering
    \includegraphics[width=0.85\linewidth]{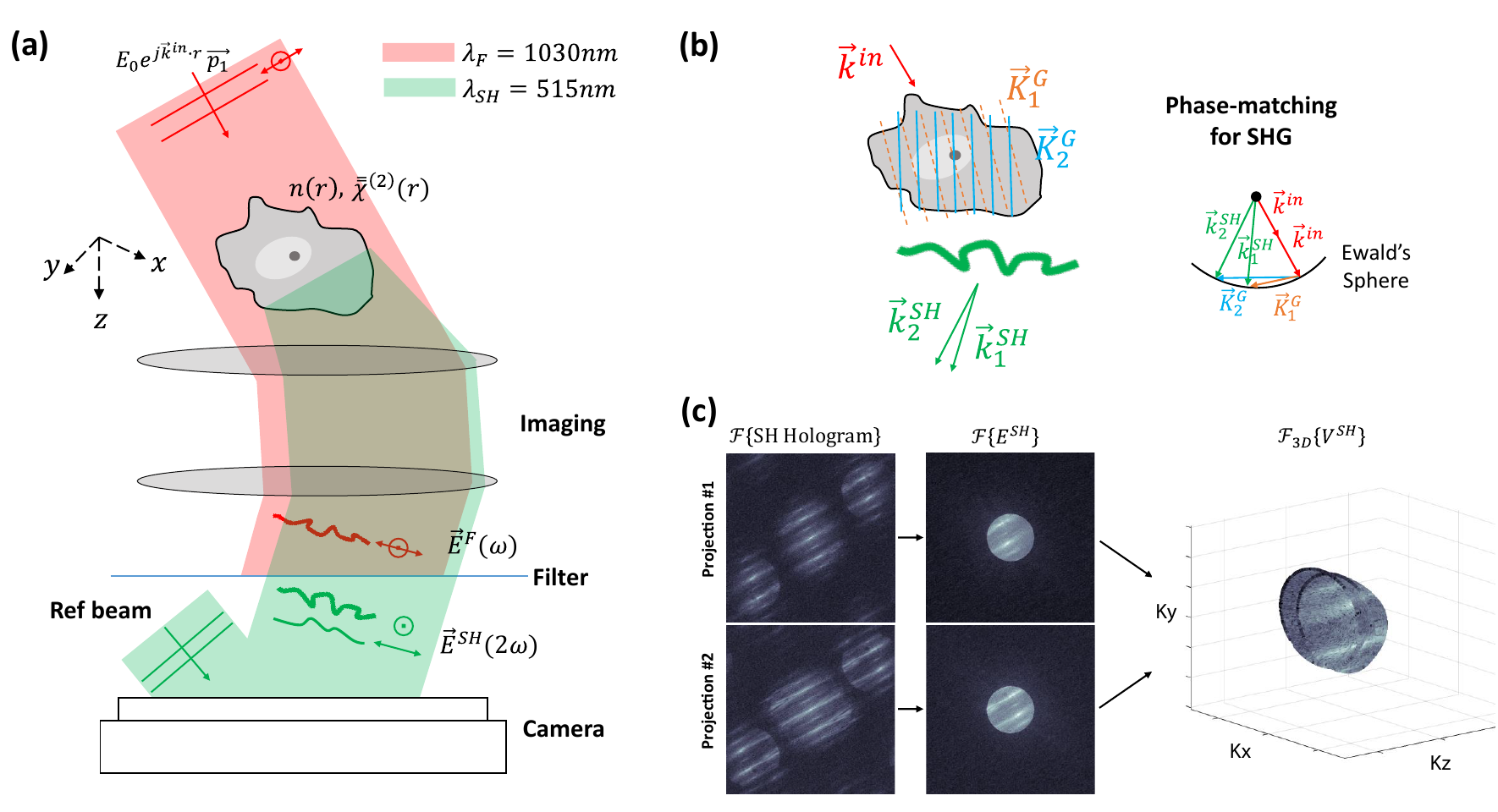}
    \caption{(a) Schematic of the SH-ODT configuration. The sample is illuminated with various input plane waves with wave vectors $\overrightarrow{k}^{in}$ and polarization states $\overrightarrow{p}_{illum}$, generating SH light (green) in two orthogonal polarization states. The SH beam is imaged with a 4F system and overlaid with a reference beam to form the SH hologram on the camera. (b) The phase-matching condition for SHG which is the basis of the Fourier diffraction theorem. (c) Filling the Fourier domain of the SH scattering potential using the complex SH fields in the Fourier domain.} 
    \label{fig:schematic}
\end{figure}

The core concept behind SH-ODT involves illuminating a sample with an inhomogeneous linear refractive index, $n(r)$, and an inhomogeneous second-order susceptibility, $\chi^{2}(r)$ from different angles with a high-power coherent light. The SH-generated beam is imaged on a camera using a 4F system and overlapped with an off-axis reference beam to create a complex interference pattern, as illustrated in Fig.~\ref{fig:schematic}a. Using SH-holography, the complex SH beams are measured for different illumination angles, and processed to reconstruct the 3D $\chi^{2}(r)$ distribution. As discussed in the introduction, the second order susceptibility differs from zero only in the regions where the crystalline structure of the sample is non-centrosymmetric. The up-conversion process of the two photons at the fundamental wavelength, $\lambda_f$, to a photon at half the wavelength $\lambda_{SH}=\lambda_f/2$ through a virtual state transition is described by two coupled Helmoltz equations \cite{boyd2020nonlinear}. Due to the tensorial nature of $\chi^{2}(r)$, a fully vectorial formalism must be considered to take into account the polarization of light in the nonlinear Helmholtz equation. 

For weakly scattering objects illuminated by a plane wave with wavevector $\overrightarrow{k}^{in}$ we can use the first-order Born approximation at the fundamental frequency and neglect the re-scattering process of the generated SH field $\overrightarrow{E}^{SH}(r)$. As more exhaustively detailed in the Supplementary Material, under these assumptions we can write

\begin{equation}
    \overrightarrow{E}^{SH}(r) = \int{G^{SH}(\mathbf{r}-\mathbf{r}') \cdot E_0^2 e^{2j\vec{k}^{in}\cdot \mathbf{r}'} \cdot \bar{\bar{V}}^{SH}(\mathbf{r}') \overrightarrow{Q}_{illum} d\mathbf{r}'}
    \label{eqn:SHGreenHelmoltzSolution}
\end{equation}

where $G^{SH}(\mathbf{r}-\mathbf{r}')=e^{jk_{SH}n_0 |r-r'|}/{ | \mathbf{r}-\mathbf{r}' |}$ is the Green's function of the SH Helmholtz equation with $k_{SH}=2\pi/\lambda_{SH}$, $\bar{\bar{V}}^{SH}(\mathbf{r}')=\frac{k_{SH}^{2}}{4\pi}\bar{\bar{\chi}}^{(2)}(\mathbf{r}')$ is the SH scattering potential tensor with the size $3\times6$, and we define the second-order vector $\overrightarrow{Q}_{illum} = [p_{x}^2, p_{y}^2, p_{z}^2, 2p_{y}p_{z}, 2p_{x}p_{z}, 2p_{x}p_{y}]^T$. The vector $\vec{p}_{illum}=[p_x,p_y,p_z]^T$ denotes the polarization state of the incident beam as $\overrightarrow{E}^I(\mathbf{r}') = E_0e^{j\vec{k}^{in}\cdot \mathbf{r}'}\vec{p}_{illum}$ and is always perpendicular to the illumination wave-vector, $\vec{p}_{illum} \cdot \vec{k}^{in} = 0$.

Next, we can invert Eq.~\ref{eqn:SHGreenHelmoltzSolution} using the Fourier diffraction theorem. We discuss the scalar and vectorial inversion in detail in the Methods and supplementary document. For each illumination angle, we vary the input polarization state $\vec{p}_{illum}$ and measure the vectorial components of the generated SH. Since the $\chi^{(2)}$ is a rank-3 tensor, we can have 3 independent experiments or polarization states for each illumination angle, in contrast to the linear case in which we can only have two independent states \cite{saba2021polarization}. The theoretical details of the polarization-sensitive SH-ODT are presented in section 2 of the Supplementary document. We can see that a $2 \times 3$ subset of the SH scattering potential tensor can be approximately found as \cite{yu2022optical},

\begin{equation}
\bar{\bar{V}}_{2 \times 3}^{SH}(K_x-2k_x^{in},K_y-2k_y^{in},K_z-2k_z^{in}) =\\ \frac{K_z}{2{\pi}jE_0^2}\mathcal{F}_{2D}\left\{ \begin{bmatrix}
E^{SH}_{x1} &E^{SH}_{x2}  &E^{SH}_{x3} \\
E^{SH}_{y1} &E^{SH}_{y2}  &E^{SH}_{y3} 
\end{bmatrix}\left [\vec{s}_1,\vec{s}_2,\vec{s}_3\right ]^{-1} \right\}\left(K_x,K_y\right)
\label{eqn:SH-ODT_Vectorial}
\end{equation}
where $\mathcal{F}_{2D}$ is the 2D Fourier transform, $K_x$ and $K_y$ are the Fourier components in the transverse direction, and $K_z = \sqrt{(k_{SH}n_0^{2\omega})^2-K_x^2-K_y^2}$, $E^{SH}_{ij}$ is the 2D SH complex field at the imaging plane polarized along $i \in \left \{ x,y \right \}$ for the input polarization state $j \in \left \{ 1,2,3 \right \}$. $\vec{s}_{j} = [p_{x}^2, p_{y}^2, 2p_{x}p_{y}]^T$ is a part of the vector $\vec{Q}_{illum}$ for the input polarization state $j$, and the $\bar{\bar{V}}^{SH}_{2 \times 3}$ is a matrix containing 6 elements of the full SH scattering potential tensor as,

\begin{equation}
\bar{\bar{V}}^{SH}_{2 \times 3} = \frac{k_{SH}^2}{4\pi} \begin{bmatrix}
\chi^{(2)}_{11} &\chi^{(2)}_{12}  &\chi^{(2)}_{16} \\ 
\chi^{(2)}_{21} &\chi^{(2)}_{22}  &\chi^{(2)}_{26} 
\end{bmatrix}
\label{eqn:V_tensor}
\end{equation} 

The phase-matching condition in the generation of the SH signal is a key point implicit in Eq.~\ref{eqn:SH-ODT_Vectorial}. We can describe the second-order susceptibility of the sample in the Fourier domain by decomposing its spatial distribution into several periodic gratings, each represented with a spatial frequency vector $\overrightarrow{K}^G$. If we also decompose the generated SH beam into several plane waves, each represented with a wave-vector $\overrightarrow{K}^{SH} = \left[K_x,K_y,K_z \right]^T$, the phase matching condition implies that $ \overrightarrow{K}^G= \overrightarrow{K}^{SH}-2\overrightarrow{k}^{in}$. This phase-matching condition, which is visually depicted in Fig.~\ref{fig:schematic}b, can be derived from Eq.~\ref{eqn:SH-ODT_Vectorial}: the different spatial frequency components of the SH scattering potential, corresponding to different gratings, are related to the Fourier frequencies of the SH field following the conditions imposed by phase-matching.

Eq.~\ref{eqn:SH-ODT_Vectorial} shows how the complex SH-generated fields for 3 different independent input polarization states can be combined to create the Ewald's spheres and fill the 3D Fourier domain of the SH scattering potential tensor. The reconstruction of a scalar or an effective $\chi^{(2)}$ distribution can be achieved by simply rewriting Eq.~\ref{eqn:SH-ODT_Vectorial} in a scalar formalism. This is discussed in detail in section 2 of the supplementary document. The process of filling the Fourier domain of the SH scattering potential is shown in Fig.1 c. After holographically measuring the complex SH fields, we can use Eq.~\ref{eqn:SH-ODT_Vectorial} to fill the 3D Fourier domain of each element of the tensor, and we can take the inverse 3D Fourier transform for each element to calculate its 3D spatial distribution. Based on this relationship, the Ewald's spheres at SH are two-folded bigger than those at the fundamental frequency and their centers shifted of $2\vec{k}_{in}$.

\begin{figure}[t]
    \centering
    \includegraphics[width=0.99\linewidth]{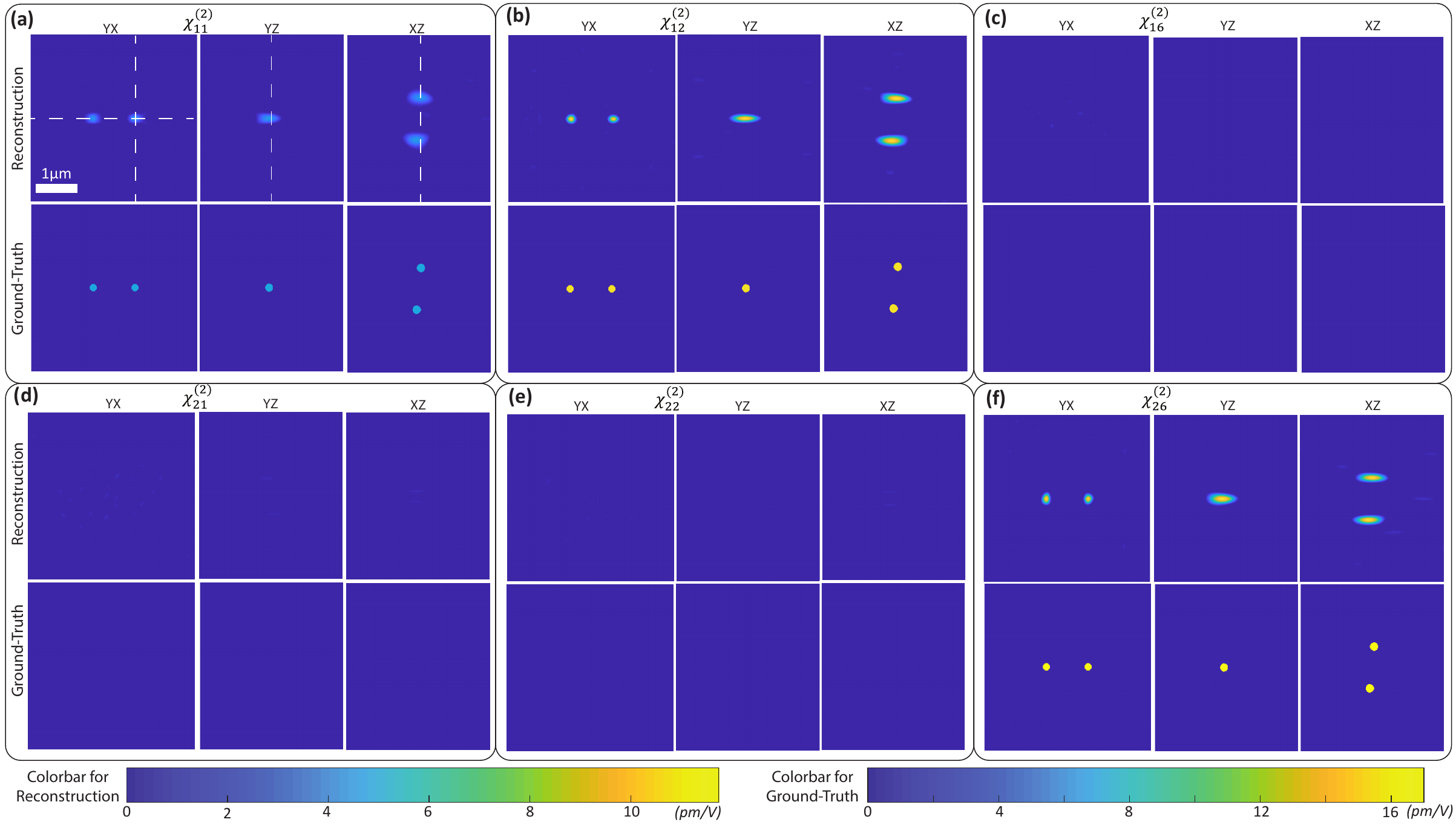}
    \caption{TV-based iterative reconstruction of second-order susceptibility tensor using synthetic data. The 6 elements of the second-order susceptibility tensor achieved using Eq.~\ref{eqn:TV} are presented in (a) $\chi^{(2)}_{11}$, (b) $\chi^{(2)}_{12}$, (c) $\chi^{(2)}_{16}$, (d) $\chi^{(2)}_{21}$, (e) $\chi^{(2)}_{22}$, and (f) $\chi^{(2)}_{26}$, respectively.} 
    \label{fig:simulationTV}
\end{figure}

\subsection{3D $\chi^{(2)}$ reconstruction based on synthetic data}\label{sec21}
In order to validate our reconstruction method, we numerically modeled the SHG from a 3D distribution of Barium titanate (BTO) nano-particles immersed in a background medium with the refractive index of $n_0 = 1.7$. BTO nano-particles have a $4mm$ group symmetry with non-zero $\chi^{(2)}_{15}$, $\chi^{(2)}_{24} = \chi^{(2)}_{15}$, $\chi^{(2)}_{31}$, $\chi^{(2)}_{32} = \chi^{(2)}_{31}$, and $\chi^{(2)}_{33}$ elements in their second-order susceptibility tensor. This scenario is interesting as it represents a scientifically relevant example which can be numerically modeled with reasonable computational costs. We generate complex fields at fundamental, $\lambda = 1030nm$, and SH, $\lambda = 515 nm$ modeled with two frequency-domain Finite-element-method (FEM) simulations, at each of these wavelengths, coupled to each other. The simulations are performed with a commercial FEM solver (COMSOL Multiphysics) and are elaborated in Methods. We illuminate the sample with different angles and collect the complex fields at the fundamental and SH wavelengths. We use these data for the 3D reconstruction of our sample. The sample consists of two BTO nano-particles with a diameter of $200nm$ aligned with the same crystalline orientation, i.e. the cz-axis oriented along the X-axis of the experimental coordinate system. 
The particles are illuminated with a circular pattern and maximum angle from the normal direction of $10^{\circ}$. We have 3 input polarization states: X-polarized, Y-polarized, and diagonally-polarized ($45^{\circ}$) illumination. The SH-generated light is calculated for all the angles and input polarization states. 

We use the synthetic data for different input polarization states and different illumination angles calculated using FEM and use first-order Born approximation and Eq.~\ref{eqn:SH-ODT_Vectorial} to calculate the 3D distribution of the SH susceptibility. 

Similarly to the linear case, the direct reconstruction of the SH scattering potential is underestimated and elongated along the optical axis due to the missing frequency components. We can use an iterative ODT reconstruction with a piece of prior knowledge about the sample to improve the reconstruction. An iterative reconstruction for SH-ODT based on a Total-variation (TV) regularization is implemented and used to improve the reconstruction fidelity. This iterative approach is detailed in Methods. 

Fig.~\ref{fig:simulationTV} shows the comparison between the iterative reconstruction based on sparsity regularizer and the ground truth for all the 6 elements of the tensor. Although Born approximation and the missing cone problem result in a slight underestimation of the nonlinear susceptibility values, the comparison between the two confirms that the reconstruction method returns an accurate prediction of the 3D nonlinear tensor distribution and with zero $\chi^{(2)}_{16}$, $\chi^{(2)}_{21}$, and $\chi^{(2)}_{22}$ elements, as the crystalline orientation implies.


\subsection{Experimental results on BTO nano-particle}

\begin{figure}[t]
    \centering
    \includegraphics[width=0.99\linewidth]{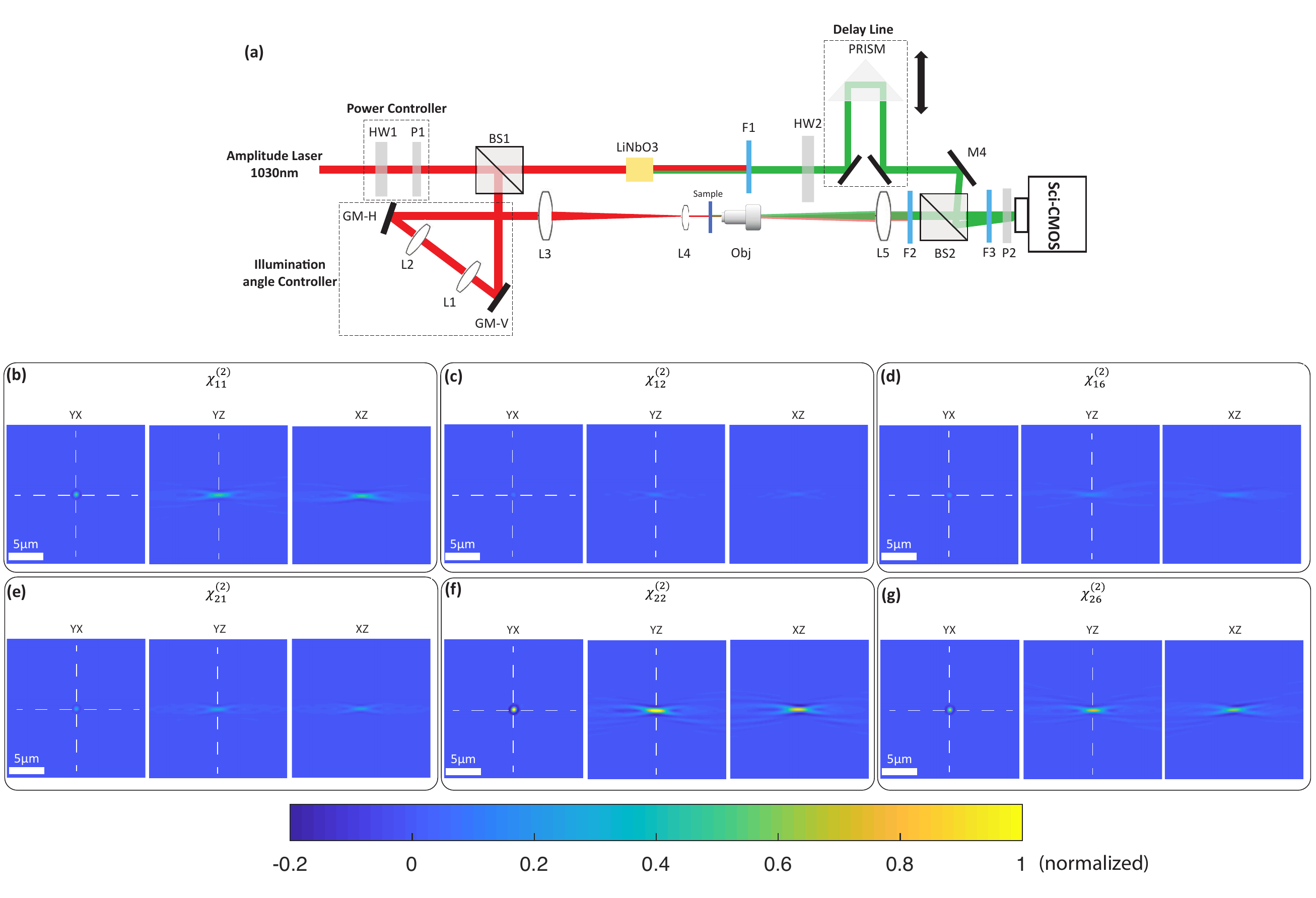}
    \caption{(a) Experimental setup for multi-angle SH-holography. (b-g) 3D reconstruction of second-order susceptibility tensor for a 200nm BTO nano-particle. The 6 elements of the second-order susceptibility tensor achieved using Eq.~\ref{eqn:SH-ODT_Vectorial} are presented in (b) $\chi^{(2)}_{11}$, (c) $\chi^{(2)}_{12}$, (d) $\chi^{(2)}_{16}$, (e) $\chi^{(2)}_{21}$, (f) $\chi^{(2)}_{22}$, and (g) $\chi^{(2)}_{26}$, respectively. Each figure presents the 3D reconstruction in YX, YZ, and XZ planes.} 
    \label{fig:BTO_experiment}
\end{figure}

The 2D measurement of complex SHG fields for different illumination directions requires a multi-angle SH-holographic setup. Such an experimental system is depicted in Fig.~\ref{fig:BTO_experiment}a and it is discussed thoroughly in Methods. In an SH-ODT setup which consists of a signal and a reference path, the sample is illuminated with a polarized high-power coherent light from different illumination angles. The illumination angles are controlled with a pair of Galvo mirrors. A set of filters and a polarizer are used to remove the fundamental beam and set the polarization of the measured SH light. On the reference path, a uniform SH beam is generated using a nonlinear crystal and it is combined with the signal SH beam to form the hologram on the camera. 

After capturing the digital hologram on the camera, a Fourier domain hologram extraction is used (as shown in Fig.~\ref{fig:schematic}c) to retrieve the phase and the amplitude of the SH-generated beam. 

As proof of concept of our method, we first demonstrate our experimental setup by reconstructing the 3D $\chi^{2}$ tensor of a BTO particle immersed in water. BTO has a larger refractive index in comparison with water, however Born-approximation is still valid due to the small size of the nano-particle (diameter = 200nm) \cite{chen1998validity}. We collect fundamental and SH holograms by illuminating the sample from 180 different angles and extracting the phase and amplitude of the SH emission and fundamental scattering for each projection and polarization. The experiment is done using three input polarization states, linear X-polarized, linear Y-polarized, and linear $45^{\deg}$-polarized. These three experiments correspond to $\vec{s}_1$, $\vec{s}_2$, and $\vec{s}_3$ vectors where the polarization vector $\vec{p}_i$ can be calculated for every illumination angle and for each input polarization state $i$ using the rotation matrix introduced in our previous work \cite{saba2021polarization}. Thus, we can calculate the $\vec{s}_i$ vector based on its definition in Sec.~\ref{sec:1}. For each input polarization, we measure the two orthogonal output polarization states, X-polarized and Y-polarized. Having the holographic measurements of the phase and amplitude of the SH fields, the 3D tomographic reconstruction of the $\chi^{2}$ tensor is shown . In this way the six elements of the $\chi^{2}$ tensor can be reconstructed and the result is presented in Fig.~\ref{fig:BTO_experiment}b. It should be mentioned that the reconstructions are calculated using the SH fields and there is no pre-knowledge about the orientation of the crystalline axis of the nano-particle in this experiment. However, having the group symmetry of BTO, we can find the crystalline orientation of the reconstructed nano-particle.

\subsection{Experimental results on muscle tissue}
\begin{figure}[t]
    \centering
    \includegraphics[width=0.90\linewidth]{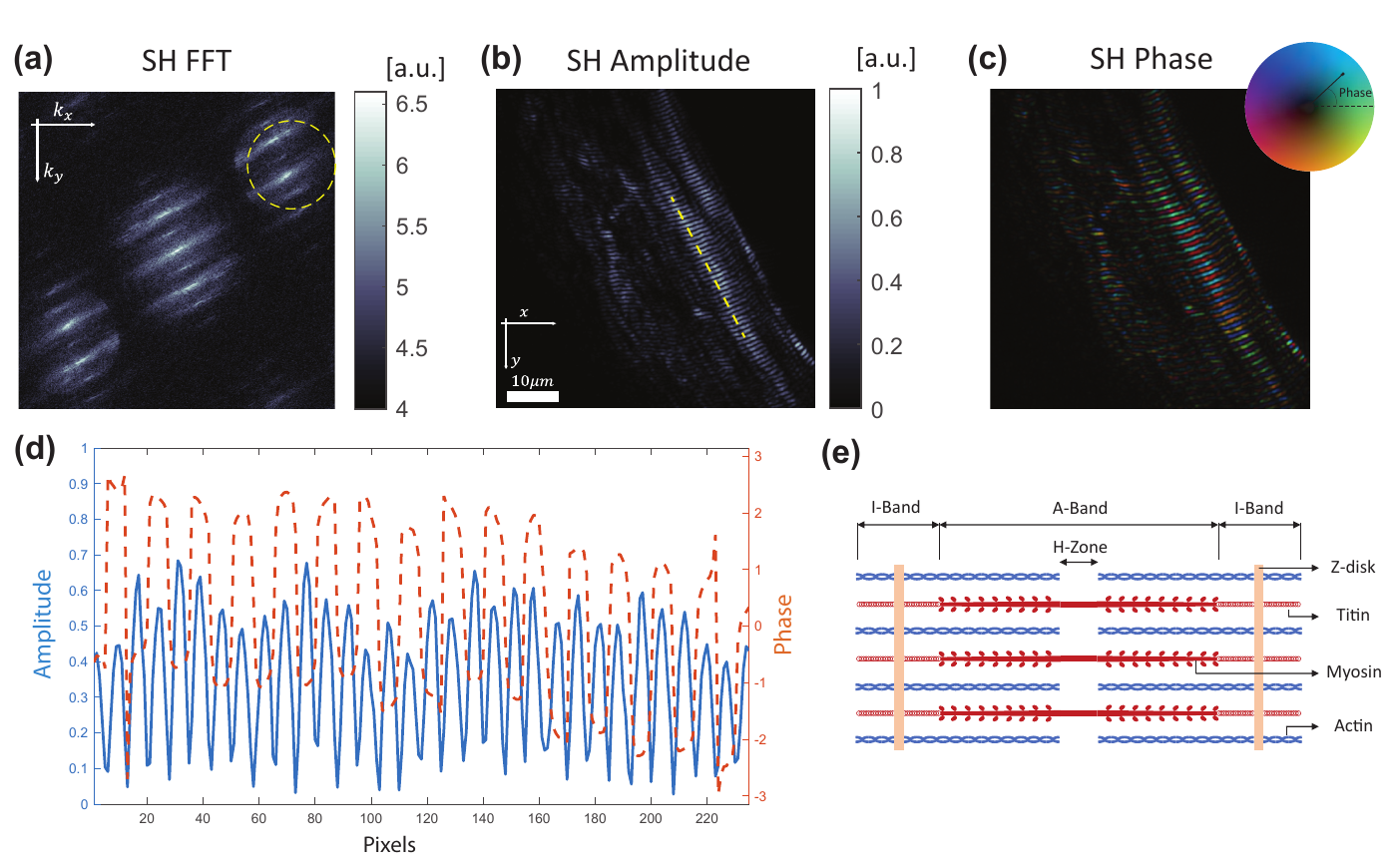}
    \caption{(a) Fourier transform of the SH hologram from muscle fiber. (b-c) Holographic extraction of the amplitude and phase of the SH-generated field from the muscle fiber as the sample illuminated with a tilted fundamental Gaussian wave. The amplitude values are normalized and the phase map is shown in the color code of (c) as the brightness is modulated with the amplitude. (d) The 1D profile of the amplitude (blue) and phase (orange) of the SH-generated field along the dashed line is shown in (b). We can see the periodic amplitude and phase variation, amplitude dips between myosin crystals, and $\pm \pi$ jumps in phase. (e) Structure of the muscle tissue sarcomere.}
    \label{fig:chap4_fig3}
\end{figure}
As mentioned in the introduction, SH-microscopy is a useful label-free and background-free imaging modality for biological samples that include fibrous proteins with endogenous second-order susceptibility, such as collagen and
myosin. This technique enables to study the structure of the muscle fibers, and retrieve useful microscopic morphological information about the supramolecular structures of myosin\cite{plotnikov2006characterization,nucciotti2010probing, dubreuil2018polarization}. In this perspective, we investigated the viability of our 3D SH-ODT technique to study muscle tissue fibers. We used a \textit{ex vivo} cryo-embedded mouse muscle with a 20 µm thickness, fixed with paraformaldehyde (PFA), immersed in water as the background medium, and sandwiched between two \#1 cover-slips.
\begin{figure}[t]
    \centering
    \includegraphics[width=0.95\linewidth]{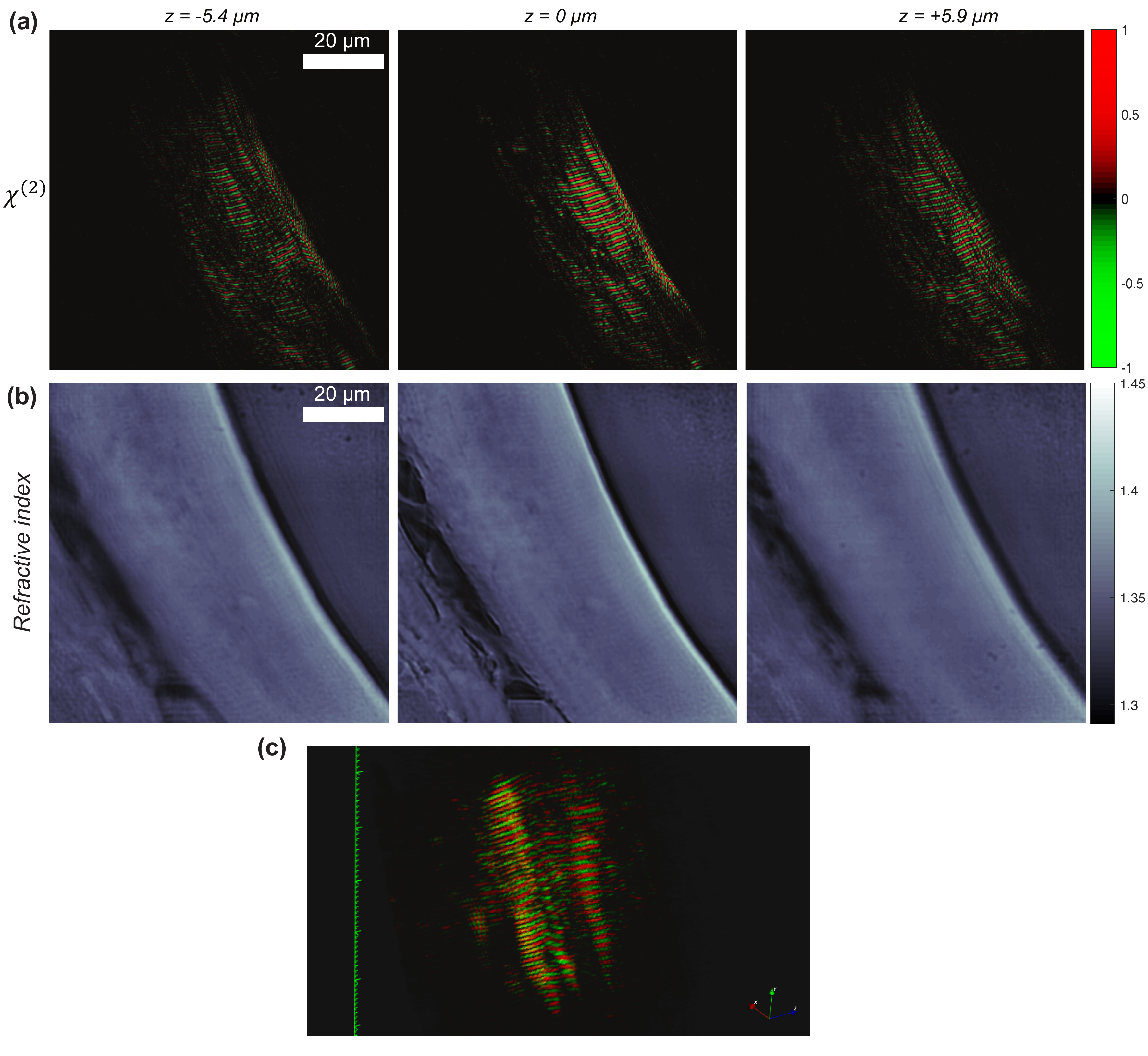}
    \caption{SH Tomographic reconstruction. (a) 2D YX profile of the 3D reconstruction of $\chi^{(2)}$ distribution in three different Z planes. The values are normalized to the [-1,1] range. (b) 2D YX profile the 3D refractive index reconstruction in three different Z planes. (c) 3D rendering of the $\chi^{(2)}$ reconstruction using SH-ODT. A 3D rendering video can be found in the supplementary media 1.}
    \label{fig:muscle_ODT}
\end{figure}

Fig.~\ref{fig:chap4_fig3}a-c reports the SH holographic acquisitions from by a skeletal muscle fiber. The 2D Fourier transform of the SH hologram generated by the interference of the endogenous SHG signal from muscle fiber and the reference beam is shown in Fig.~\ref{fig:chap4_fig3}a. We use Fourier-domain holographic extraction to achieve the amplitude and phase of the SH emission from the muscle fiber as illuminated with a tilted fundamental beam. The amplitude and phase of the complex SH field are shown in Fig.~\ref{fig:chap4_fig3}b-c. The fiber structure is composed of repeating connective units, known as sarcomeres, sketched in Fig.~\ref{fig:chap4_fig3}e, which are arranged periodically along the fiber direction. The periodic structure of sarcomeres can be clearly seen in the amplitude of the SH-emission in Fig.~\ref{fig:chap4_fig3}b. We take a closer look at this periodic structure by plotting the profile of the amplitude and phase of the SH emission on the dashed line shown in Fig.~\ref{fig:chap4_fig3}b. The phase and amplitude plots are presented in Fig.~\ref{fig:chap4_fig3}d. We see that the phase of the SH emission has ${\pm \pi}$ jumps when the amplitude drops to zero. This observation implies that within a period of the sarcomere, the myosin filaments are mirrored with respect to the H-zone, as clearly demonstrated in Fig.~\ref{fig:chap4_fig3}e. When the crystal is mirrored, the $\chi^{(2)}$ values change their sign, which leads to a $\pi$ shift in the SH phase profile. The same effect was also observed by interferometric SH-microscopy \cite{rivard2013imaging}. It should be noted that the centro-symmetry implies that the amplitude of the SH emission vanishes at the ends of the sarcomere unit, known as z-disks, and also at the middle of these periodic units, which is manifested in the dips of the amplitude profile of Fig.~\ref{fig:chap4_fig3}d.

We collected holograms by illuminating the sample from 180 different angles and extracting the phase and amplitude of the SH emission and fundamental scattering for each projection. As the target muscle fiber is mainly oriented along the y-direction, the dominant component in the first column of its $\chi^{(2)}$-tensor is $\chi^{(2)}_{y1}$ \cite{nucciotti2010probing}. As a consequence, in this case, we consider the x-polarized illumination and measure the y-polarized complex SH-emission to reconstruct the 3D distribution of the $-21-$ element of the tensor.

Fig.~\ref{fig:muscle_ODT}a shows the yx profile of the $\chi^{(2)}$ reconstruction in three different z-planes. This figure clearly highlights the structure of myosin proteins which are periodically arranged. A 3D reconstruction of the linear refractive index distribution is presented in Fig.~\ref{fig:muscle_ODT}b using Rytov approximation \cite{devaney1981inverse} with the holographic measurements of the linear scattering at the fundamental wavelength. A visualization of the 3D renderings of the reconstructed $\chi^{(2)}$ distribution are shown in Fig.~\ref{fig:muscle_ODT}c.

\section{Discussion}\label{sec3}
\subsection{SH-ODT Reconstruction}
The numerical analysis on BTO nanoparticles leverages on the full tensorial nature of the second harmonic generation process. Due to the complex form of the material nonlinear susceptibility, we can get different values of the $\chi^{(2)}$ tensor depending on the orientation of the crystal. The comparison between the reconstruction result and the ground truth in Fig.~\ref{fig:simulationTV} shows that our algorithm returns the correct tensorial form of the second-order susceptibility. To quantify that, we also define a mean-squared-error (MSE) reconstruction metric based on the mean value of the squared value of the difference between the reconstructed $\chi^{(2)}$ and the ground truth. The averaging is done over all of the tensorial components and voxels. For the digital phantoms represented in Fig.~\ref{fig:simulationTV}, the MSE error is $0.12$ considering the TV-based iterative reconstruction as opposed to $0.68$ for the direct reconstruction without regularization reported in Fig.~S1 in the supplementary. The iterative reconstruction can clearly improve the performance at the expense of the required computation time for performing an iterative optimization task.

In general we need to record polarization-sensitive SH-holograms and measure the complex fields for various input polarization states in order to reconstruct the tensorial form of the second-order susceptibility. However, if we assume that the sample has an approximately uniform crystalline coordinate system, and also small illumination angles, we can also extract the scalar form of Eq.~\ref{eqn:SH-ODT_Vectorial} to reconstruct a 3D distribution of an effective second-order susceptibility, $\chi^{(2)}_{eff}(r)$. For the case of muscle tissue of Fig.~\ref{fig:chap4_fig3} whose muscle fiber is mainly oriented along the y-direction, we can approximately assume that $\chi^{(2)}_{11} \approx \chi^{(2)}_{12} \approx \chi^{(2)}_{26} \approx 0$, and $\chi^{(2)}_{16} = \chi^{(2)}_{21} \neq 0, \chi^{(2)}_{22} \neq 0$. In this case, if we have an x-polarized illumination and measure the y-component of the SH beam, the scalar Fourier diffraction theorem gives us an approximated value of $\chi^{(2)}_{21}$. This is represented as SH-ODT results in Fig.~\ref{fig:muscle_ODT}. It should be noted that due to these approximations and also to the fact that the muscle fiber is not completely oriented along the y-direction, the reconstructed $\chi^{(2)}(r)$ of muscle tissue is not exactly representing the quantitative values of the -21- element of the tensor.

\subsection{SH: a Background-free modality for ODT}
The tomographic reconstruction of the BTO nano-particles with synthetic data and muscle tissue with experimental data in section~\ref{sec:2} bring up some important observations. Firstly, the SH-holographic measurements in Fig.~\ref{fig:chap4_fig3} provide very interesting features in the amplitude and phase of the SH signal emitted from the muscle tissue. The $\pm \pi$ phase jumps present details about the morphology of the myosin proteins. On the other hand, the SH amplitude informs about the thickness of the myosin proteins. Furthermore, this imaging modality provides a label-free and background-free 3D image. Comparison of the reconstructed second-order susceptibility and refractive index distributions in  Fig.~\ref{fig:muscle_ODT} shows SH-ODT provides important and interesting features that cannot be seen in the linear ODT. The periodic arrangement of the sarcomeres units can be hardly seen in the linear refractive index reconstruction but with a low resolution and contrast. At variance, the second-order susceptibility reconstruction presents a detailed image of the muscle tissue fibers.

Due to the shorter wavelength of the SH signal, the resolution of the SH-ODT is twice as good as the conventional ODT, both in the transverse and axial directions. The resolution can be further improved by sample rotation or wavelength scanning. This improved resolution and contrast makes SH-ODT a  useful technique for the reconstruction of biological and non-biological samples containing non-centrosymmetric materials. These advantages of SH-ODT come at the expense of a complicated experimental setup and the high-power illumination beam. In the case of our experimental setup presented in Fig.~\ref{fig:chap4_fig3}, we use a light source with $3.5 \mathrm{\mu J}$ pulse energy and use 50\% of this power to illuminate the sample. Increasing the intensity of the illumination more than in our case could cause several problems such as bubble generation and damaging the imaging objective lens. The required exposure time to capture a high signal-to-noise ratio SH image with our illumination intensity was about $100 \mathrm{ms}$. 

\subsection{Generalization to other nonlinear processes}
We can generalize the methodology of this paper to other optical nonlinear processes rather than SHG. The reconstructed scattering potential can be written based on the phase-matching condition for the corresponding nonlinear process. As an example, for non-centrosymmetric materials, we can observe sum-frequency generation if we have two high power sources. In sum-frequency generation, two coherent sources at frequencies $\omega_1$ and $\omega_2$ combine and generate a third frequency, $\omega_1+\omega_2 \rightarrow \omega_3$. The nonlinear polarizability for this process will be,\begin{equation}
P_i (r,\omega_3)= \epsilon_0
\chi^{(2)} (r)
E_1(r,\omega_1) 
E_2(r,\omega_2)
\label{eq:SFG}
\end{equation}
where $E(r,\omega_1)$, and $E(r,\omega_2)$ represent the electric field in two pump frequencies. Considering the phase matching condition for this nonlinear process, we can write the tensorial Wolf transform  to find the sum-frequency scattering potential,
\begin{equation}
V^{SF}(K_x-k_x^{in,1}-k_x^{in,2},K_y-k_y^{in,1}-k_y^{in,2},K_z-k_z^{in,1}-k_z^{in,2}) = \frac{K_z}{2{\pi}jE_1 E_2}\mathcal{F}_{2D}\left\{E^{SF}\right\}\left(K_x,K_y\right)
\label{eqn:SF-ODT_Scalar}
\end{equation}
in which $K_x$ and $K_y$ are the Fourier components in the transverse direction, $E_1$ and $E_2$ are the amplitude of the fields at frequencies $\omega_1$, and $\omega_2$, and $K_z = \sqrt{(k_{SF}n_0^{\omega_3})^2-K_x^2-K_y^2}$. Eq.~\ref{eqn:SF-ODT_Scalar} shows how the spatial frequency domain of the scattering potential should be filled based on the 2D Fourier transform of the generated nonlinear signal. According to this equation, the phase matching condition is the key point in understanding how different frequencies of the scattering potential contribute to the 2D Fourier transform of the generated signal. If we decompose the generated signal into plane waves in the Fourier domain, the phase-matching condition should be satisfied for the plane wave of $E^{SF}\left(K_x,K_y \right)$. The phase-matching condition implies that $\overrightarrow{K} = \left [K_x,K_y,K_z \right]^T = \overrightarrow{K}_g + \overrightarrow{k}^{in,1}+\overrightarrow{k}^{in,2}$ where $\overrightarrow{K}_g$ is the frequency vector of the corresponding grating in the scattering potential which generates this plane wave of $E^{SF}$. Considering Eq.~\ref{eqn:SF-ODT_Scalar}, we can clearly see this phase-matching condition in the way that the frequency components of the scattering potential are filled. As a result, we can rewrite this equation for any parametric nonlinear conversion, such as third-harmonic generation or four-wave mixing by taking the phase-matching condition into account for that nonlinear conversion process. 

\section{Conclusion} 
Concluding, we presented ODT for the second-harmonic conversion and showed 3D reconstructions of the second-order susceptibility of the sample using multiple-angle holographic measurements of SHG. The presented formalism was based on applying the Fourier diffraction theorem for the SH Helmholtz equation. 

We developed SH-ODT for reconstructing the second-order susceptibility based on the first-order Born approximation. The proposed methodology is tested with synthetic examples achieved using FEM. We applied this technique to a biological sample, a muscle tissue fiber, and reconstructed its second-order susceptibility showing various features such as myosin crystals in the sarcomere units of the tissue. The presented SH-ODT provides high-resolution label- and background-free 3D images of details that cannot be revealed in linear ODT or QPI.

\section{Methods}

\subsection{Iterative reconstruction}
Due to the limited numerical aperture of the imaging optics, the limited number of projections, and also having the transmission configuration, we cannot retrieve all the 3D spatial frequencies of the SH scattering potential tensor elements. This issue is the same as the linear  ODT, known as the missing cone problem, which was under intensive research study in the previous years \cite{lim2015comparative}. The missing cone problem causes the 3D reconstructed scattering potential to be underestimated and also elongated along the optical axis, indicated here as the $z$-axis. One way to fix this issue is to use a piece of prior information on the reconstruction elements of the SH scattering potential tensor. Here, we use a Total-Variation (TV) sparsity condition to regularize the reconstruction of the SH inverse problem. For this purpose, after reconstructing the 3D Fourier domain of the SH scattering potential tensor using Eq.~\ref{eqn:SH-ODT_Vectorial} which we call the $mn$ element of its tensor as $\tilde{V}^{SH,recon}_{mn}\left(K_x,K_y,K_z\right)$, we optimize the following minimization task to find the $V^{SH,opt}_{mn}$,
\begin{equation}
V^{SH,opt}_{mn} = \underset{x}{\arg\min}\left[\|\mathcal{F}_{3D}\left\{x \right\}-\tilde{V}^{SH,recon}_{mn}\|^2_2+\mathcal{R}_{TV}\{x\}\right]
\label{eqn:TV}
\end{equation}
in which $\mathcal{R}_{TV}\{x\}$ is the gradient-based TV regularization on the 3D distribution of $x$. 

\subsection{Experimental setup}
The first experiments for digital harmonic holography were presented in \cite{pu2008harmonic}. We use the same principles to build a polarization-sensitive and multi-angle SH-holography setup. The polarization-sensitive holographic SH-ODT system we use to acquire our experimental SH complex fields for different samples of interest is shown in Fig~\ref{fig:chap4_fig3}a. The light source is a yttrium-doped fiber laser (Amplitude Laser Satsuma) with a wavelength of $\lambda_f=1030nm$, 280fs pulses, and a repetition rate of 125kHz. The power of the source beam is controlled with a Half-waveplate (HW1) and a polarizer (P1) which sets its polarization to the horizontal direction ($x$-axis). The source beam is split into the signal and the reference arms using a beam splitter (BS1). The polarization of the signal arm is controlled with another half-waveplate (HW2) and we can set it to 3 independent states of horizontal, vertical, and $45$-deg input polarization states. Using a galvo mirror (GM-V), we change the illumination angle in the vertical direction. GM-V is then imaged on a secondary galvo mirror (GM-H) using a 4F system (L1 and L2 lenses). GM-H controls the illumination beam angle in the horizontal direction and is imaged on the sample using another 4F system consisting of L3 and L4 lenses. This 4F system has a demagnification of 8 which makes an angular magnification with the same amount. The sample generates the SH light (green beam in Fig.~\ref{fig:chap4_fig3}a) which is imaged using a microscope objective and L5 on the Scientific CMOS camera (Andor Neo 5). On the reference side, an SH Gaussian beam is generated with a lithium-niobate LiNbO3 uniform crystal. Due to the extremely short coherent length of the light source corresponding to 280fs, the optical paths between the reference and signal arms should match. To have that, we use a delay path in the reference arm on a motorized stage to control the optical path of the reference arm. The polarization of the reference SH arm is set to 45-deg in the $xy$ lab system with HW3 to be able to get the interference with both polarization components of the SH signal beam. The reference and signal arms are combined in an off-axis configuration and their polarization is filtered with P2 in order to get the hologram of the desired output polarization. We consider two output polarization of $x$ and $y$ orientations. 

Filtering the SH light is an important issue in the SH holographic setup. In order to completely block the high-power fundamental beam we use a set of spectral filters with the total optical density of OD19. On the reference side, filter F1 drops the intensity of the fundamental beam and also gives us the ability to have some part of the fundamental beam in the case of measurement of $E^F$ and $E^I$ which is possible by removing F3. On the signal side, we have two filters, F2 and F3 which reduce the intensity of the fundamental beam. For the imaging, we use a 60X dry objective (Nikon 60X-PF Plan Fluorite, 0.85 NA, 0.31-40mm WD). The objective is dry but the sample is immersed in water. This issue causes aberrations which can be corrected numerically as we have access to the complex fields holographically. 

The desired sample of the study is illuminated with a conical illumination pattern with a maximum angle of $\theta \approx 12 ^{\circ}$ and the phase and the amplitude of the light at the SH and fundamental wavelengths are collected for different pairs of input and output polarization. We use this data for the tomographic 3D reconstruction of $\chi^{(2)}$ distribution which we discuss in the next section.

\subsection{Frequency domain FEM for SH simulations}
The reconstruction model presented in section \ref{sec:2} with synthetic data generated with finite element model using the commercial software COMSOL Multiphysics. In undepleted-pump approximation, the second harmonic generation poblem can be broken down in two consecutive processes in the frequency domain: the first one with an external plane wave excitation at fundamental frequency which returns the total field $\overrightarrow{E}^{F}(\omega)$ within the sample volume and combined with the quadratic susceptibility of the material $\chi^{(2)}$ enables to compute the second order polarizability $\overrightarrow{P}^{SH}(2\omega)=\varepsilon_0\chi^{(2)}\overrightarrow{E}^{F}(\omega)\otimes\overrightarrow{E}^{F}(\omega)$; the second one at the second harmonic with a localized current source $\overrightarrow{J}^{SH}(2\omega)=-2i\omega\overrightarrow{P}^{SH}(2\omega)$ which returns the total field $\overrightarrow{E}^{SH}$ at $2\omega$. We modeled the experimental conical excitation with 36 plane waves with wavelength $\lambda_{F}=1030$ nm matching our Yb-doped fiber laser. As a realistic benchmark case, we considered two barium titanate spheres with refractive index  $n=2.3$ and diameter $d=200$ nm embedded in a medium with index 1.7 which enable to simultaneously have a reasonable simulation volume and a weakly-scattering object far from Mie-resonances. The total fields $\overrightarrow{E}^{F}$ and $\overrightarrow{E}^{SH}$ are interpolated on a plane far from the center of mass of the two spheres and back-propagated to the sample plane through beam propagation method to undo diffraction and model the role of the imaging system. 

\bibliographystyle{ieeetr}
\bibliography{references}  

\begin{thebibliography}{10}

\bibitem{wolf1969three}
E.~Wolf, ``Three-dimensional structure determination of semi-transparent
  objects from holographic data,'' {\em Optics communications}, vol.~1, no.~4,
  pp.~153--156, 1969.

\bibitem{sung2009optical}
Y.~Sung, W.~Choi, C.~Fang-Yen, K.~Badizadegan, R.~R. Dasari, and M.~S. Feld,
  ``Optical diffraction tomography for high resolution live cell imaging,''
  {\em Optics express}, vol.~17, no.~1, pp.~266--277, 2009.

\bibitem{jin2017tomographic}
D.~Jin, R.~Zhou, Z.~Yaqoob, and P.~T. So, ``Tomographic phase microscopy:
  principles and applications in bioimaging,'' {\em JOSA B}, vol.~34, no.~5,
  pp.~B64--B77, 2017.

\bibitem{park2018quantitative}
Y.~Park, C.~Depeursinge, and G.~Popescu, ``Quantitative phase imaging in
  biomedicine,'' {\em Nature photonics}, vol.~12, no.~10, pp.~578--589, 2018.

\bibitem{lim2015comparative}
J.~Lim, K.~Lee, K.~H. Jin, S.~Shin, S.~Lee, Y.~Park, and J.~C. Ye,
  ``Comparative study of iterative reconstruction algorithms for missing cone
  problems in optical diffraction tomography,'' {\em Optics express}, vol.~23,
  no.~13, pp.~16933--16948, 2015.

\bibitem{kamilov2015learning}
U.~S. Kamilov, I.~N. Papadopoulos, M.~H. Shoreh, A.~Goy, C.~Vonesch, M.~Unser,
  and D.~Psaltis, ``Learning approach to optical tomography,'' {\em Optica},
  vol.~2, no.~6, pp.~517--522, 2015.

\bibitem{tian20153d}
L.~Tian and L.~Waller, ``3d intensity and phase imaging from light field
  measurements in an led array microscope,'' {\em optica}, vol.~2, no.~2,
  pp.~104--111, 2015.

\bibitem{saba2022physics}
A.~Saba, C.~Gigli, A.~B. Ayoub, and D.~Psaltis, ``Physics-informed neural
  networks for diffraction tomography,'' {\em Advanced Photonics}, vol.~4,
  no.~6, p.~066001, 2022.

\bibitem{saba2021polarization}
A.~Saba, J.~Lim, A.~B. Ayoub, E.~E. Antoine, and D.~Psaltis,
  ``Polarization-sensitive optical diffraction tomography,'' {\em Optica},
  vol.~8, no.~3, pp.~402--408, 2021.

\bibitem{shin2022tomographic}
S.~Shin, J.~Eun, S.~S. Lee, C.~Lee, H.~Hugonnet, D.~K. Yoon, S.-H. Kim,
  J.~Jeong, and Y.~Park, ``Tomographic measurement of dielectric tensors at
  optical frequency,'' {\em Nature Materials}, vol.~21, no.~3, pp.~317--324,
  2022.

\bibitem{boyd2020nonlinear}
R.~W. Boyd, {\em Nonlinear optics}.
\newblock Academic press, 2020.

\bibitem{freund1986second}
I.~Freund and M.~Deutsch, ``Second-harmonic microscopy of biological tissue,''
  {\em Optics letters}, vol.~11, no.~2, pp.~94--96, 1986.

\bibitem{stoller2003quantitative}
P.~Stoller, P.~M. Celliers, K.~M. Reiser, and A.~M. Rubenchik, ``Quantitative
  second-harmonic generation microscopy in collagen,'' {\em Applied optics},
  vol.~42, no.~25, pp.~5209--5219, 2003.

\bibitem{plotnikov2006characterization}
S.~V. Plotnikov, A.~C. Millard, P.~J. Campagnola, and W.~A. Mohler,
  ``Characterization of the myosin-based source for second-harmonic generation
  from muscle sarcomeres,'' {\em Biophysical journal}, vol.~90, no.~2,
  pp.~693--703, 2006.

\bibitem{nucciotti2010probing}
V.~Nucciotti, C.~Stringari, L.~Sacconi, F.~Vanzi, L.~Fusi, M.~Linari,
  G.~Piazzesi, V.~Lombardi, and F.~Pavone, ``Probing myosin structural
  conformation in vivo by second-harmonic generation microscopy,'' {\em
  Proceedings of the National Academy of Sciences}, vol.~107, no.~17,
  pp.~7763--7768, 2010.

\bibitem{dubreuil2018polarization}
M.~Dubreuil, F.~Tissier, L.~Le~Roy, J.-P. Pennec, S.~Rivet, M.-A.
  Giroux-Metges, and Y.~Le~Grand, ``Polarization-resolved second harmonic
  microscopy of skeletal muscle in sepsis,'' {\em Biomedical optics express},
  vol.~9, no.~12, pp.~6350--6358, 2018.

\bibitem{pu2008harmonic}
Y.~Pu, M.~Centurion, and D.~Psaltis, ``Harmonic holography: a new holographic
  principle,'' {\em Applied Optics}, vol.~47, no.~4, pp.~A103--A110, 2008.

\bibitem{masihzadeh2010label}
O.~Masihzadeh, P.~Schlup, and R.~A. Bartels, ``Label-free second harmonic
  generation holographic microscopy of biological specimens,'' {\em Optics
  express}, vol.~18, no.~10, pp.~9840--9851, 2010.

\bibitem{shaffer2011second}
E.~Shaffer, P.~Marquet, and C.~Depeursinge, ``Second harmonic phase microscopy
  of collagen fibers,'' in {\em Multiphoton Microscopy in the Biomedical
  Sciences XI}, vol.~7903, pp.~63--68, SPIE, 2011.

\bibitem{hu2020harmonic}
C.~Hu, J.~J. Field, V.~Kelkar, B.~Chiang, K.~Wernsing, K.~C. Toussaint, R.~A.
  Bartels, and G.~Popescu, ``Harmonic optical tomography of nonlinear
  structures,'' {\em Nature Photonics}, vol.~14, no.~9, pp.~564--569, 2020.

\bibitem{yu2022optical}
W.~Yu, X.~Li, B.~Wang, J.~Qu, and L.~Liu, ``Optical diffraction tomography of
  second-order nonlinear structures in weak scattering media: theoretical
  analysis and experimental consideration,'' {\em Optics Express}, vol.~30,
  no.~25, pp.~45724--45737, 2022.

\bibitem{chen1998validity}
B.~Chen and J.~J. Stamnes, ``Validity of diffraction tomography based on the
  first born and the first rytov approximations,'' {\em Applied optics},
  vol.~37, no.~14, pp.~2996--3006, 1998.

\bibitem{rivard2013imaging}
M.~Rivard, C.-A. Couture, A.~K. Miri, M.~Lalibert{\'e}, A.~Bertrand-Grenier,
  L.~Mongeau, and F.~L{\'e}gar{\'e}, ``Imaging the bipolarity of myosin
  filaments with interferometric second harmonic generation microscopy,'' {\em
  Biomedical Optics Express}, vol.~4, no.~10, pp.~2078--2086, 2013.

\bibitem{devaney1981inverse}
A.~Devaney, ``Inverse-scattering theory within the rytov approximation,'' {\em
  Optics letters}, vol.~6, no.~8, pp.~374--376, 1981.

\end{thebibliography}


\begin{thebibliography}{1}
\newcommand{\enquote}[1]{``#1''}

\bibitem{saba2021polarization}
A.~Saba, J.~Lim, A.~B. Ayoub, E.~E. Antoine, and D.~Psaltis,
  \enquote{Polarization-sensitive optical diffraction tomography,}
  {\protect\JournalTitle{Optica}} \textbf{8}, 402--408 (2021).

\end{thebibliography}






\end{document}


\title{Second-harmonic optical diffraction tomography: supplemental document}
\author{} 
\begin{abstract}
\end{abstract}
\maketitle

\section{Wave propagation in nonlinear inhomogeneous media}
The nonlinear polarizability for the SHG process is as, 
\begin{equation}
\begin{pmatrix}
P_x(2\omega)
\\ 
P_y(2\omega)
\\ 
P_z(2\omega)
\end{pmatrix}
= 
\epsilon_0
\bar{\bar{\chi}}^{(2)}
\begin{pmatrix}
E_x(\omega)^2\\
E_y(\omega)^2\\ 
E_z(\omega)^2\\ 
2E_y(\omega)E_z(\omega)\\ 
2E_x(\omega)E_z(\omega)\\
2E_x(\omega)E_y(\omega)
\end{pmatrix}
\label{eq:b1}
\end{equation}
where $E_x(\omega)$, $E_y(\omega)$, and $E_z(\omega)$ are the electric field components at the fundamental wavelength. Here, we introduce the notation  $\left(\overrightarrow{E}_F \star \overrightarrow{E}_F\right)$ notation to write the $6 \times 1$ squared field vector. Disregarding the depletion of the pump approximation and neglecting light depolarization resulting from high-gradient variations in the refractive index distribution, we are able to formulate two Helmholtz equations at fundamental and SH wavelengths,
\begin{multline}
    \nabla^2 \overrightarrow{E}_{F}(r,\omega) +
    k_0^2n^2(r,\omega) \overrightarrow{E}_{F}(r,\omega)  =  0
    \\
    \nabla^2 \overrightarrow{E}_{SH}(r,2\omega) +
    4k_0^2n^2(r,2\omega) \overrightarrow{E}_{SH}(r,2\omega)  = 
    -4k_0^2 \bar{\bar{\chi}}^{(2)}(r) \cdot \left(\overrightarrow{E}_F(r,\omega) \star \overrightarrow{E}_F (r,\omega)\right) 
    \qquad
\label{eq:b2} 
\end{multline}
where $\overrightarrow{E}_{F}(r,\omega)$ and $\overrightarrow{E}_{SH}(r,2\omega)$ are the field vectors at the fundamental and SH wavelengths, $k_0 = 2\pi / \lambda_F$ is the fundamental wave number, and $n(r,\omega)$ and $n(r,\omega)$ are the refractive index distributions at the fundamental and SH wavelengths. We can consider these two values equal in the case of no dispersion. It should be mentioned that this does not change the generality of the formalism and dispersion can be added if it is characterized for the background medium. The first equation of Eq.~\ref{eq:b2} describes the linear scattering behavior of the fundamental light. We can rewrite the right side of the second equation using the first-order Born approximation by replacing $\overrightarrow{E}_F(r)$ with the incident field $\overrightarrow{E}_i(r)$. 
The incoming field constitutes a plane wave propagating uniformly through the background medium, maintaining a constant amplitude and polarization state. If we regard this incident plane wave as the electric field vector, $\overrightarrow{E}_i(r) = E_0 e^{j\overrightarrow{k}_{in}\cdot r} \left[p_x, p_y, p_z \right]^{T}$, we can write the Helmholtz equation at the SH wavelength in this form,
\begin{equation}
    \nabla^2 \overrightarrow{E}_{SH}(r,2\omega) +
    4k_0^2n^2(r,2\omega) \overrightarrow{E}_{SH}(r,2\omega)  = 
    4 \pi E_0^2 e^{2j\overrightarrow{k}_{in}\cdot r} \bar{\bar{V}}^{(SH)}(r) \cdot \overrightarrow{Q}_{illum}
\label{eq:b3} 
\end{equation}
where $\overrightarrow{Q}_{illum} = \left[p_x, p_y, p_z \right]^{T} 
\star \left[p_x, p_y, p_z \right]^{T} = \left[p_{x}^2, p_{y}^2, p_{z}^2, 2p_{y}p_{z}, 2p_{x}p_{z}, 2p_{x}p_{y}\right]^T$
is the $6 \times 1$ squared polarization vector, and $\bar{\bar{V}}^{(SH)} = k_0^2/\pi \bar{\bar{\chi}}^{(2)}(r)$ is the SH scattering potential. Eq.~\ref{eq:b3} is the vectorial Helmholtz equation governing the SH field in a second-order nonlinear medium, characterized with the nonlinear scattering potential, $\bar{\bar{V}}^{(SH)}$. This equation results from three approximations so far: (i) exclusive involvement of $\omega$ and $2\omega$ frequencies in the interaction, (ii) disregard of the fundamental beam depletion, and (iii) application of the first-order Born approximation in the scattering of the fundamental light. The inhomogeneity of the sample's refractive index distribution, denoted as $n^2(r,2\omega)$, controls the scattering of the generated SH field within the sample. A fourth approximation to further simplify Eq.~\ref{eq:b3} is to neglect the SHG rescattering due to the refractive index inhomogeneity. This approximation implies to replace $n(r,2\omega)$ with $n_0(2\omega)$ in the left side of Eq.~\ref{eq:b3}. Using this approximation, we can write the integral solution of Eq.~\ref{eq:b3} using Green's function of this equation,
\begin{equation}
    \overrightarrow{E}^{SH}(r) = \int{G^{SH}(\mathbf{r}-\mathbf{r}') \cdot E_0^2 e^{2j\overrightarrow{k}_{in}\cdot \mathbf{r}'} \cdot \bar{\bar{V}}^{SH}(\mathbf{r}') \overrightarrow{Q}_{illum} d\mathbf{r}'}
\label{eq:b4}
\end{equation}
in which $G^{SH}(\mathbf{r}-\mathbf{r}')=e^{2jk_{0}n_0(2\omega) | r-r'  |}/{ | \mathbf{r}-\mathbf{r}' |}$ is the Green's function at SH wavelength. Eq.~\ref{eq:b4} is the simplest formalism for the calculation of the SHG from a sample with inhomogeneous and anisotropic second-order susceptibility. 

\section{Fourier diffraction theorem for SHG}
In the next step, our objective is to reverse the equation denoted as Eq.\ref{eq:b4} to introduce the Fourier diffraction theorem for SHG and a reconstruction technique for determining the second-order scattering potential through the utilization of SH-generated fields at multiple angles. Employing the Fourier transform of Eq.\ref{eq:b4}, and through the application of the Fourier diffraction theorem, we obtain,
\begin{equation}
\bar{\bar{V}}^{SH}(K_x-2k_x^{in},K_y-2k_y^{in},K_z-2k_z^{in}) \overrightarrow{Q}_{illum} (k_x^{in},k_y^{in},k_z^{in})= \frac{K_z e^{-jK_z z_0}}{2{\pi}jE_0^2}\mathcal{F}_{2D}\left\{\overrightarrow{E}^{SH}\right\}\left(K_x,K_y\right)
\label{eq:b5}
\end{equation}
where $K_x$, $K_y$, and $K_z$ are the spatial frequencies. It should be noted that $\bar{\bar{V}}^{SH}$ is a $3 \times 6$ tensor and $\overrightarrow{Q}_{illum}$ is a $6 \times 1$ vector. As a result, their matrix product will be a $3 \times 1 $ vector which is coherent with the size of the SH field vector. 

In the scalar field approximation, we completely neglect the tensorial form of Eq.~\ref{eq:b5} and reconstruct the scalar SH scattering potential, $V^{SH}$ using the scalar SH-generated field. As a result, we have,
\begin{equation}
{V}^{SH}(K_x-2k_x^{in},K_y-2k_y^{in},K_z-2k_z^{in}) = \frac{K_z e^{-jK_z z_0}}{2{\pi}jE_0^2}\mathcal{F}_{2D}\left\{E^{SH}\right\}\left(K_x,K_y\right)
\label{eq:b6}
\end{equation}

However, the inversion is more complicated for the tensorial second-order susceptibility. The SH scattering potential is a $3 \times 6$ matrix. As a result, we need to measure 3 components of the complex SH field vector for 6 independent states of $\overrightarrow{Q}_{illum}$ to be able to find all of the 18 elements of the SH scattering potential tensor. Nevertheless, measurement of $E^{SH}_z$ is impossible as the polarization vector is perpendicular to the cameras. Additionally, there can be only 3 independent states for the $\overrightarrow{Q}_{illum}$ of the plane wave. We consider these 3 independent states as generated by the input field polarized along $x$-axis, $y$-axis, and $+45^{\circ}$ in the xy plane. Considering the approximation that the incident illumination angle with respect to the optical axis, $z$, is small, we can use the same approximation in \cite{saba2021polarization}. As a result, we completely remove the $z$-term in Eq.~\ref{eq:b5} by considering $p_z$ as zero in this equation and removing the last row of this equation which is responsible for $E^{SH}_z$. Regarding this approximation, we can write,
\begin{equation}
\bar{\bar{{V}}}^{SH}_{2 \times 6}(K_x-2k_x^{in},K_y-2k_y^{in},K_z-2k_z^{in})\begin{pmatrix}
p_x^2 (\overrightarrow{k}^{in})\\
p_y^2(\overrightarrow{k}^{in})\\
0\\ 
0\\ 
0\\ 
2p_x(\overrightarrow{k}^{in})p_y(\overrightarrow{k}^{in})
\end{pmatrix} = \frac{K_z e^{-jK_z z_0}}{2{\pi}jE_0^2}\mathcal{F}_{2D}\left\{\begin{pmatrix}
E^{SH}_x\\ 
E^{SH}_y
\end{pmatrix}\right\}\left(K_x,K_y\right)
\label{eq:b7}
\end{equation}
where $\bar{\bar{V}}^{SH}_{2 \times 6}$ contains only the first two rows of the SH scattering potential tensor. We can see that there are three zeros in the $\overrightarrow{Q}_{illum}$ which makes the third, fourth, and fifth columns of the SH scattering potential tensor impossible to be retrieved. Thus, we remove these columns and rewrite Eq.~\ref{eq:b7} with a $2 \times 3$ scattering potential,
\begin{equation}
\bar{\bar{V}}^{SH}_{2\times 3}(K_x-2k_x^{in},K_y-2k_y^{in},K_z-2k_z^{in}) \cdot \overrightarrow{s}= \frac{K_z e^{-jK_z z_0}}{2{\pi}jE_0^2}\mathcal{F}_{2D}\left\{\overrightarrow{E}^{SH}\right\}\left(K_x,K_y\right)
\label{eq:b8}
\end{equation}
in which $\overrightarrow{s}=[p_x^2,p_y^2,2p_xp_y]^{T}$. As mentioned earlier, we can invert Eq.~\ref{eq:b8} with 3 states for the vector $\overrightarrow{s}$. Considering the 3 states of $\overrightarrow{s}$ for $x$-polarized, $x$-polarized, and $45^{\circ}$-polarized input field as $\vec{s}_1$, $\vec{s}_2$, and $\vec{s}_3$. Therefore, we can invert Eq.~\ref{eq:b8} as follows,
\begin{multline}
\bar{\bar{V}}_{2 \times 3}^{SH}(K_x-2k_x^{in},K_y-2k_y^{in},K_z-2k_z^{in}) =\\ \frac{K_ze^{-jK_z z_0}}{2{\pi}jE_0^2}\mathcal{F}_{2D}\left\{ \begin{bmatrix}
E^{SH}_{x1} &E^{SH}_{x2}  &E^{SH}_{x3} \\ 
E^{SH}_{y1} &E^{SH}_{y2}  &E^{SH}_{y3} 
\end{bmatrix}\left [\vec{s}_1,\vec{s}_2,\vec{s}_3\right ]^{-1} \right\}\left(K_x,K_y\right)
\label{eq:b9}
\end{multline}
in which $E^{SH}_{ij}$ is the complex 2D SH field at the imaging plane polarized along $i \in \left \{ x,y \right \}$ for the input polarization state of $j \in \left \{ 1,2,3 \right \}$.

\section{Fourier diffraction theorem with corrected-field Born approximation}
We can improve the calculation of the SH field by modifying the first-order Born approximation in Eq.~\ref{eq:b2}. For the sake of simplicity, we write this section in the scalar field regime but it can be easily written in the tensorial form. We can rewrite Eq.~\ref{eq:b2} as follows,
\begin{equation}
    \nabla^2 E^{SH}(r) +
    4k_0^2n^2(r) {E}^{SH}(r)  = 
    -4\pi V^{SH}(r) \left(E^F(r)/E^i(r)\right)^2 \left({E}^i(r)\right)^2
\label{eq:b10} 
\end{equation}
if we define $\psi(r) = \left(E_F(r)/E_i(r)\right)^{-2}$ and use the relationship of $\nabla^2 (fg) = f \nabla^2 g+2\nabla f \cdot \nabla g +g \nabla^2 f$, we can write,
\begin{equation}
    \nabla^2 \left( \psi(r) E^{SH}(r)\right)
    - 2 \nabla \psi(r) \cdot \nabla E^{SH}
    + \left(4k_0^2n^2(r)\psi(r)-\nabla^{2}\psi(r)\right) {E}^{SH}(r)  = 
    -4\pi V^{SH}(r) \left({E}^i(r)\right)^2
\label{eq:b11} 
\end{equation}

Using the slowly varying approximation we can assume that $\psi(r)$ and $E^{SH}$ are varying slowly. As a result, we can neglect the second term of the left side and replace $4k_0^2n^2(r)\psi(r)-\nabla^{2}\psi(r)$ with $4k_0^2n_0^2(r)\psi(r)$ in the third term of the left side in Eq.~\ref{eq:b11}. This way, we will have,
\begin{equation}
    \left\{\nabla^2 +
    4k_0^2n^2(r)\right\} \left({E}_{SH}(r) \psi(r) \right)  = 
    -4\pi V^{SH}(r) \left({E}_i(r)\right)^2
\label{eq:b12} 
\end{equation}
Comparing Eq.~\ref{eq:b12} with Eq.~\ref{eq:b2}, we can simply apply the Fourier diffraction theorem and write a similar equation to Eq.~\ref{eq:b6} as,
\begin{equation}
{V}^{SH}(K_x-2k_x^{in},K_y-2k_y^{in},K_z-2k_z^{in}) = \frac{K_z e^{-jK_z z_0}}{2{\pi}jE_0^2}\mathcal{F}_{2D}\left\{E^{SH}(r) \left(\frac{E_i(r)}{E_F(r)}\right)^2\right\}\left(K_x,K_y\right)
\label{eq:b13}
\end{equation}
Eq.~\ref{eq:b13} corrects the SH field with the fundamental field scattering and can slightly improve the first-order Born approximation considering the scattering of the fundamental beam due to the inhomogeneity of the refractive index distribution.

\section{BTO nano-particles reconstruction without TV regularization}

\begin{figure}[t]
    \centering
    \includegraphics[width=0.99\linewidth]{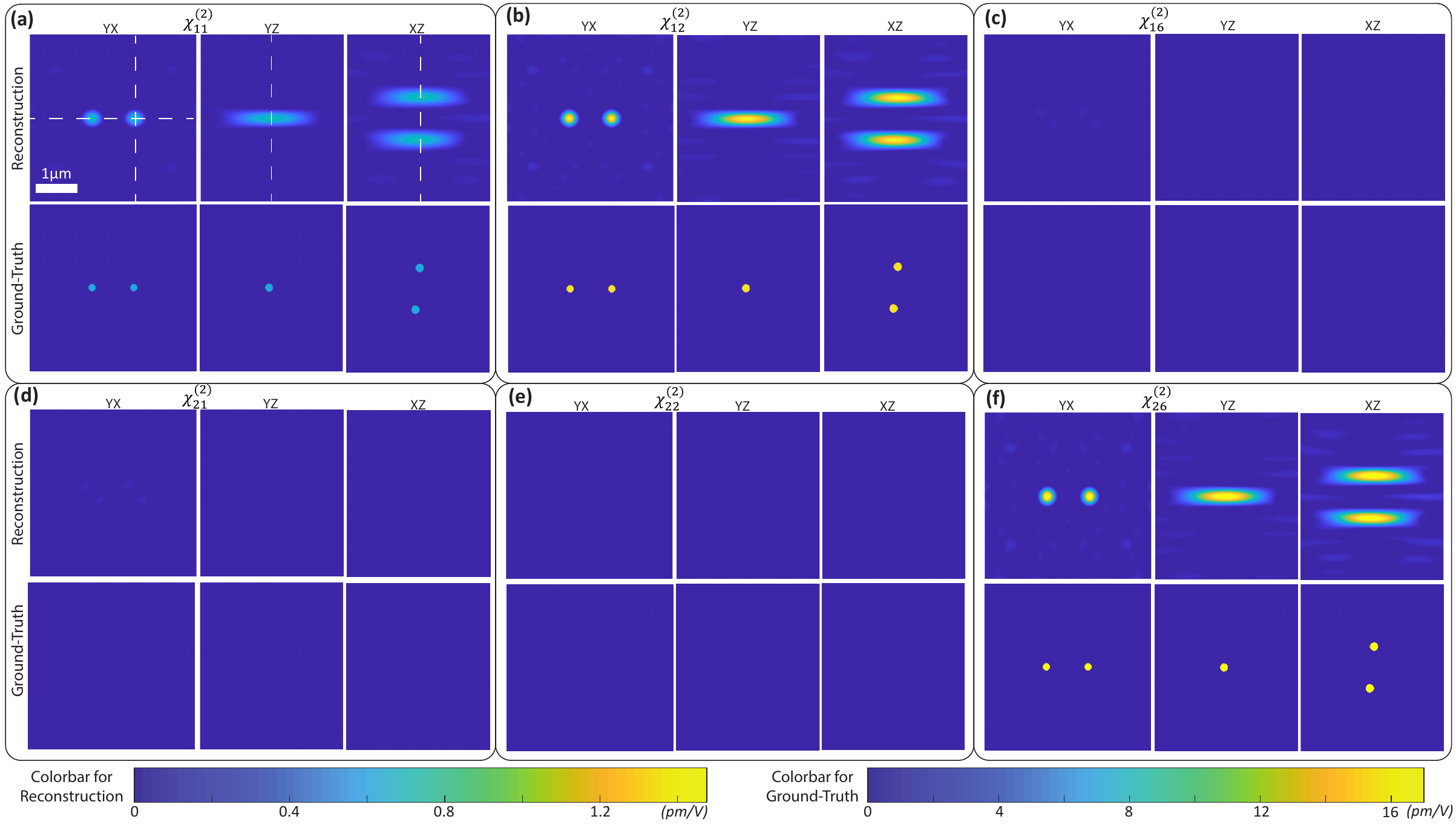}
    \caption{3D reconstruction of second-order susceptibility tensor using synthetic data. The 6 elements of the second-order susceptibility tensor achieved using Eq.~\ref{eqn:SH-ODT_Vectorial} are presented in (a) $\chi^{(2)}_{11}$, (b) $\chi^{(2)}_{12}$, (c) $\chi^{(2)}_{16}$, (d) $\chi^{(2)}_{21}$, (e) $\chi^{(2)}_{22}$, and (f) $\chi^{(2)}_{26}$, respectively. Each figure presents the 3D reconstruction in YX, YZ, and XZ planes.} 
    \label{fig:simulation}
\end{figure}

In section 2.1 of the manuscript, we formulate 3D reconstruction of the $\chi^{2}$ tensor using Fourier diffraction theorem and we further exploit regulairzed 3D reconstruction using total-variation denoising regularization to improve the axial elongation. Here, we present the original 3D reconstruction of the $\chi^{2}$ tensor of the BTO nano-particles from the synthetic data in Fig.~\ref{fig:simulation}. If we compare this with Fig. 2 of the manuscript, we can see a lot of improvement in the underestimation of the 3D reconstruction and also axial elongation. This figure shows the advantage of using total variation regularization clearly. 

\section{Muscle tissue experiment}

\begin{figure}[t]
    \centering
    \includegraphics[width=0.96\linewidth]{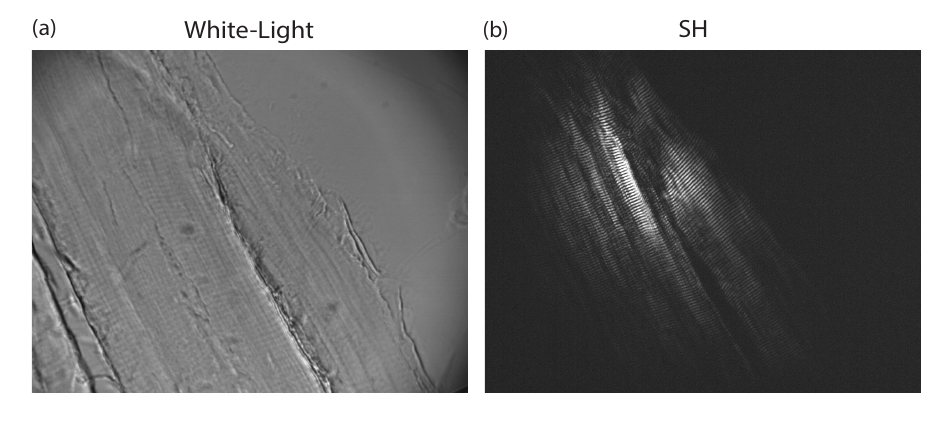}
    \caption{(a) Bright-field microscopy of the muscle fiber tissue using green light. (b) Wide-field SH image of the muscle fiber in the same region.} 
    \label{fig:muscle_BFSH}
\end{figure}

In the experimental arrangement depicted in the manuscript, a muscle tissue sample with multiple fibers was positioned. To enable Second Harmonic (SH) bright-field imaging, we obstructed the SH reference beam. Concurrently, a bright-field image of the sample was captured using a green LED source. The SH image unveils intriguing details of the sample, notably the periodic organization of sarcomere units, which remain obscured in the bright-field counterpart. The bright-field and SH images can be found in Fig.~\ref{fig:muscle_BFSH}.
\bibliography{sample}
